\documentclass[twocolumn,showpacs,prl,superscriptaddress]{revtex4}
\usepackage{graphicx}
\usepackage{color}
\usepackage{amssymb,amsfonts,amsmath}
\usepackage{natbib}

\begin{document}

\title{Multiple scale theory of topology driven pattern on directed networks}

\author{Silvia Contemori}
\affiliation{Universit\`{a} degli Studi di Firenze, Dipartimento di Matematica e Informatica, viale Morgagni 67/a, 50134 Firenze, Italia}
\author{Francesca Di Patti} \affiliation{Universit\`{a} degli Studi di Firenze, Dipartimento di Fisica e Astronomia, CSDC and INFN, via G. Sansone 1, 50019 Sesto Fiorentino, Italia}
\author{Duccio Fanelli} \affiliation{Universit\`{a} degli Studi di Firenze, Dipartimento di Fisica e Astronomia, CSDC and INFN, via G. Sansone 1, 50019 Sesto Fiorentino, Italia}
\author{Filippo Miele} \affiliation{Universit\`{a} degli Studi di Firenze, Dipartimento di Fisica e Astronomia, via G. Sansone 1, 50019 Sesto Fiorentino, Italia}

\date{\today} 


\begin{abstract}
Dynamical processes on networks are currently being considered in different domains of cross-disciplinary interest. Reaction-diffusion systems hosted on directed graphs are in particular relevant for their widespread applications, from  neuroscience, to computer networks and traffic systems. Due to the peculiar spectrum of the discrete Laplacian operator, homogeneous fixed points can turn unstable, on a directed support, because of the topology of the network, a phenomenon which cannot be induced on undirected graphs. A linear analysis can be performed to single out the conditions that underly the instability. The complete characterization of the patterns, which are eventually attained beyond the linear regime of exponential growth, calls instead for a full non linear treatment. By performing a multiple time scale perturbative calculation, we here derive an effective equation for the non linear evolution of the amplitude of the most unstable mode, close to the threshold of criticality. This is a Stuart-Landau equation whose complex coefficients appear to depend on the topological features of the embedding directed graph. The theory proves adequate versus simulations, as confirmed by operating with a paradigmatic reaction-diffusion model. 
\end{abstract}


\maketitle

Networks are undoubtedly gaining considerable importance in the modeling of natural and artificial phenomena \cite{latora, boccaletti}. They define in fact the natural playground for a large plethora of problems, that assume a heterogeneous support for the connections among constituents. In the brain, for instance, neuronal networks provide the skeleton for the efficient transport of the electric signal \cite{wyller07}. The crowded world of cells in general is shaped by veritable routes, the microtubules, that result in an intricate cobweb of interlinked paths \cite{lodish}. The flow of information on Internet, and its multifaceted applications, heavily rely on the topology of the underlying, global and local, network of contacts. Human mobility patterns, with their consequences for transportation design and epidemic control, configure, at a plausible level of abstraction, as effective graphs, linking different spatial locations.

Reactions occur on each node between species that populate the examined system. Individual actors (molecules, humans, cars or even bits of information) can relocate to distant sites, when exploring the network on which they are physically confined. This latter process is ruled by diffusion on the heterogeneous, network-like support, different avenues of transport being available to the microscopic entities, as dictated by the adjacency matrix associated to the hosting graph. The non trivial interplay between reactions and diffusion can instigate the emergence of spatially extended motifs \cite{othmer1, othmer2}, which reflect the inherent ability of the system to spontaneously self-organize and consequently perform dedicated tasks. In general, when space reduces to a regular lattice or a symmetric graph, the dynamics is uniquely responsible for the onset of the instability which eventually materializes in the observed macroscopic and collective patterns. These are, for instance, the celebrated Turing patterns that, in recent years, have received much attention also in light of their applicability on networks \cite{nakao2010, asllani2013, dipatti}. 

In applications, however, networks are not always symmetric, or, undirected, as customarily termed. Often a connection between adjacent nodes imposes a specific direction to the journey, thus resulting in a so called directed graph.

The map of neural connection is manifestly asymmetric, because of the neurons' physiology \cite{kandel}. In connectome models in fact the coarse-grained maps of the brain reveal an asymmetric arrangements of connections at different spatial scales \cite{sporns, lichman}. Cytoskeletal molecular motors move unidirectionally along an oriented polymer tracks. The cyberworld is also characterized by an asymmetric routing of the links \cite{john2010}. As traffic is concerned, several routes can be crossed in one direction only, thus breaking the symmetry between pairs of nodes. When reaction-diffusion systems are considered on directed networks, topology does matter. Surprisingly, patterns can rise on a directed support, even if they are formally impeded on a regular, continuum or discrete, spatial medium. The mathematics of this process has been recently investigated in \cite{asllani14}, where the conditions for the instability are obtained in the framework of a standard linear analysis calculation. The patterns which manifest as a byproduct of the aforementioned instability, reflect however the nonlinearities which are accommodated for into the model and that are, by definition, omitted in the linear analysis theory. In other words, the conditions for the emergence of topology driven patterns for a reaction diffusion system on a directed graph can be singled out, but the characterization of the subsequent non linear stage of evolution solely relies on numerical methods.      

In this paper we aim at filling this gap, by analytically deriving an effective equation for the evolution of the amplitude of the unstable mode, near the threshold of criticality. The spatial characteristics of the generated patterns owe to the spectrum of the Laplacian operator that governs the diffusion process. The analysis builds on a multiple time scale treatment originally devised in \cite{kuramoto75, kuramotoBook, kuramoto74, newell}, and recently reconsidered for the rather specific case of a reaction-diffusion system placed on top of a symmetric network and subject to weak couplings \cite{nakao2014}. At variance, we here focus on the case of a directed graph and assume arbitrary large diffusion coefficients. This latter condition results in a complexification of the analytical procedure: the linear calculation is carried out in a $N$ dimensional space, $N$ being the number of nodes in the graphs. The extension to non-linear orders proves consequently more demanding. A Stuart-Landau (SL) equation is eventually derived for the amplitude of the unstable mode. Unprecedently, the coefficients of the SL equation reflect the topology of the network,  the factual drive to the instability. Simulations performed for the Brusselator model, a reaction-diffusion system of pedagogical relevance, confirm the predictive adequacy of the analytical solution, obtained in the framework of the effective SL scenario.
\section{Results}
Consider a directed network composed of $N$ nodes. The topological structure of the network is encoded in the asymmetric adjacency matrix, here denoted by ${\bf A}$. The element $A_{ij}$ is equal to $1$, if nodes $i$ and $j$ are connected, or $0$ otherwise.  Each node $i$ is populated by two species, whose concentrations are respectively labeled $x_i$ and $y_i$. The species may react or diffuse throughout the network, as specified by the following general set of equations
\begin{equation}
\begin{aligned}
\frac{d}{dt} x_i =& f(x_i,y_i, \boldsymbol{ \mu})+D_x \sum_{j=1}^N \Delta_{ij} x_j  \\
\frac{d}{dt} y_i = &g(x_i,y_i,  \boldsymbol{ \mu})+D_y \sum_{j=1}^N \Delta_{ij}  y_j 
\end{aligned}
\label{eq:systemMF}
\end{equation}
where $f(\cdot, \cdot, \boldsymbol{ \mu})$ and $g(\cdot, \cdot, \boldsymbol{ \mu})$ are nonlinear functions of the concentration, which descend from the specific reactions being at play. $\boldsymbol{ \mu}$ is a vector of arbitrary dimension, where we imagine stored the scalar parameters, as e.g. the rates associated to the reactions chain, which appear to modulate the process of mutual and self-interaction. ${\bf \Delta}$ stands for the Laplacian matrix associated to the examined network. More explicitly, $\Delta_{ij}=A_{ij}-\delta_{ij}k_i$ where $k_i=\sum_{j}A_{ij}$ represents the degree of node $i$. $D_x$ and $D_y$ are the diffusion coefficients. To make contact with the analysis carried out in \cite{asllani14}, we shall deal with perfectly balanced networks, namely graphs characterized by an identical number of ingoing and outgoing links. We will then assume that the equations \eqref{eq:systemMF} admit a homogeneous stable equilibrium identified as $(x^*, y^*)$.  To save notations, it is convenient to define a vector which contains the concentrations $x_i$ and $y_i$ at any node location $i=1,..,N$, namely $\boldsymbol{ \mathsf{x}}=\left ( x_1, \dots, x_N, y_1, \ldots, y_N \right )^T$. Consequently, $\boldsymbol{ \mathsf{x}}^*$ will refer to the aforesaid steady state. We are here interested in the conditions that yield a destabilization of the homogeneous stationary stable solution $\boldsymbol{ \mathsf{x}}^*$, as follows the injection of a tiny perturbation which activates non trivial interferences between diffusion and reaction terms. As anticipated above, the directed spatial support matters: it can actively seed an instability, which is instead prevented to occur when the problem is formulated on a symmetric spatial backing. In the following, we shall briefly recall the main steps of the linear analysis theory: these are in fact propedeutic to the forthcoming developments, which aim at the full non linear picture. 

\subsection{Linear stability analysis}
Introduce a small inhomogeneous perturbation, $\delta x_i$ and $\delta y_i$, to the uniform steady state. In formulae, $(x_i, y_i)=(x^*, y^*)+(\delta x_i, \delta y_i)$ for $i=1, \ldots, N$. Substitute the latter ansatz into equations \eqref{eq:systemMF}: Taylor expanding the obtained system and packing  $\delta x_i$ and $\delta y_i$ into the column vector ${\bf u}=\left ( \delta x_1, \dots, \delta x_N, \delta y_1, \ldots, \delta y_N \right )^T$,  one immediately finds the following equation for the time evolution of  ${\bf u}$: 
\begin{equation}\label{eq:equation_u}
\frac{\partial}{dt} { \bf u= (  L+ D)  u} +  \mathcal{M}  { \bf u u } +  \mathcal{N}  { \bf u u u}
\end{equation}
where ${\bf L}$ and ${\bf D}$ are two $2N \times 2N$ block matrices 
\begin{equation*}
{ \bf L} = \left (
\begin{array}{c|c}
f_{x} (\boldsymbol{ \mathsf{x}}^*) \mathbb{I}_{N}  & f_y (\boldsymbol{ \mathsf{x}}^*)  \mathbb{I}_{N} \\
\phantom{i } & \phantom{i }\\
\hline
\phantom{i } & \phantom{i }\\
g_x (\boldsymbol{ \mathsf{x}}^*)  \mathbb{I}_{N}  & g_y (\boldsymbol{ \mathsf{x}}^*)  \mathbb{I}_{N} 
\end{array}
\right) 
\quad 
{ \bf D } = \left (
\begin{array}{c|c}
D_x  {\bf  \Delta }  & \mathbb{O}_{N} \\
\phantom{i } & \phantom{i }\\
\hline
\phantom{i } & \phantom{i }\\
\mathbb{O}_{N}  & D_y {\bf  \Delta }
\end{array}
\right) 
\end{equation*}
with $ \mathbb{I}_{N} $ and $\mathbb{O}_{N}$ denoting, respectively,  the identity matrix and the null matrix of size $N$.  $\mathcal{M}  { \bf u u }  $ and $ \mathcal{N}  { \bf u u u}$ are  symbolic notations, mutuated from \cite{kuramotoBook}. These are vectors whose $i$th  components respectively read
\begin{equation*}
\begin{aligned}
\left ( \mathcal{M}  { \bf u u }  \right  )_i &=\frac{1}{2 !}
\begin{cases}
\displaystyle \sum_{j,k \in \{i, i+N \} } \frac{\partial^2 f(\boldsymbol{ \mathsf{x}}^*)}{\partial \mathsf{x}_j \partial \mathsf{x}_k} u_j u_k  & \text{for  } i \leqslant N \\
\displaystyle\sum_{j,k \in \{i, i-N \} }  \frac{\partial^2 g(\boldsymbol{ \mathsf{x}}^*)}{\partial \mathsf{x}_j \partial \mathsf{x}_k} u_j u_k   & \text{for  } i > N
\end{cases}
 \\
\left ( \mathcal{N}  { \bf u u u }  \right  )_i & =\frac{1}{3 !}
\begin{cases}
\displaystyle \sum_{j,k,l \in \{i, i+N \} } \frac{\partial^3 f(\boldsymbol{ \mathsf{x}}^*)}{\partial \mathsf{x}_j \partial \mathsf{x}_k \partial \mathsf{x}_l} u_j u_k u_l & \text{for  } i \leqslant N \\
\displaystyle\sum_{j,k,l \in \{i, i-N \} }  \frac{\partial^3 g(\boldsymbol{ \mathsf{x}}^*)}{\partial \mathsf{x}_j \partial \mathsf{x}_k \partial \mathsf{x}_l } u_j u_k u_l   & \text{for  } i > N
\end{cases}
\end{aligned}
\end{equation*}
The study of the stability of $(x^*,y^*)$ relies on the linear part of equation \eqref{eq:equation_u}
\begin{equation}\label{eq:eigLinear}
\left ( {\bf L} + {\bf D} \right ) {\bf u} = \lambda {\bf u}
\end{equation}

To solve the above linear system, one needs to introduce the eigenvalues $\Lambda ^{(\alpha)}$ and eigenvectors $\boldsymbol{\phi}^{(\alpha)}$ of the Laplacian operator  \cite{nakao2010,asllani14}. These are
solutions of the eigenvalue problem ${\bf \Delta} \boldsymbol{\phi}^{(\alpha)}= \Lambda^{(\alpha)} \boldsymbol{\phi}^{(\alpha)}$ for $\alpha = 1, \ldots, N $. Importantly, when the hosting network is directed, the eigenvalues of the Laplacian are complex. This latter property is ultimately responsible for the peculiar behavior of reaction-diffusion systems placed on asymmetric graphs, as compared to their undirected homologues.  The inhomogeneous perturbations $\delta x_i$ and $\delta y_i$ can be expanded as:
\begin{equation}\label{eq:expansionPerturb}
\delta x_i = \sum_{\alpha=1}^N c_{\alpha} e  ^{\lambda^{(\alpha)} t} \phi_i^{\alpha} \quad 
\delta y_i = \sum_{\alpha=1}^N \eta^{(\alpha)} c_{\alpha}  e^{\lambda^{(\alpha)} t} \phi_i^{\alpha} 
\end{equation}
where $c_{\alpha}$ depend on initial conditions, and $\eta^{(\alpha)}$ will be self-consistently specified later on. By inserting  \eqref{eq:expansionPerturb} into \eqref{eq:eigLinear}, yields  $N$ copy of the following system
\begin{equation}\label{eq:system2}
\left ( 
\begin{array}{c c }
\hspace{-0.2cm}f_x + D_x \Lambda^{(\alpha)} -\lambda^{(\alpha)}& \hspace{-0.1cm} f_y\hspace{-0.2cm}\\
\hspace{-0.2cm} \phantom{.} &  \hspace{-0.1cm}  \phantom{.}  \hspace{-0.2cm}\\
\hspace{-0.2cm}g_x & \hspace{-0.1cm} g_y+ D_y \Lambda^{(\alpha)}   -\lambda^{(\alpha)} \hspace{-0.2cm}
\end{array}
\right)  \left ( 
\begin{array}{c}
\hspace{-0.15cm}1\hspace{-0.15cm}\\
\hspace{-0.15cm}\phantom{.}\hspace{-0.15cm}\\
\hspace{-0.15cm}\eta^{(\alpha)}\hspace{-0.15cm}
\end{array}
\right)
=
\left (\begin{array}{c}
\hspace{-0.1cm}0\hspace{-0.1cm}\\
\hspace{-0.1cm}\phantom{.}\hspace{-0.1cm}\\
\hspace{-0.1cm}0\hspace{-0.1cm}
\end{array}\right)
\end{equation}
which admits a non trivial solution provided
\begin{equation}\label{eq:dispersion}
\det \left ( 
\begin{array}{c c }
f_x + D_x \Lambda^{(\alpha)} -\lambda^{(\alpha)}& f_y\\
 \phantom{.} &  \phantom{.} \\
g_x & g_y+ D_y \Lambda^{(\alpha)}   -\lambda^{(\alpha)} 
\end{array}
\right) =0 \qquad .
\end{equation}

Equation \eqref{eq:dispersion} returns a second order polynomial for $\lambda^{(\alpha)}$  as a function of $\Lambda^{(\alpha)}$, known as the dispersion relation. The stability of $(x^*, y^*)$ depends on the sign of the real part of $\lambda^{(\alpha)}$, here termed $\lambda^{(\alpha)}_{Re}$: if $\lambda^{(\alpha)}_{Re}$ is negative $\forall \alpha$, the $(x^*, y^*)$  is stable, while it turns unstable if  $\lambda^{(\alpha)}_{Re}$ crosses punctually the x-axis. In this case, the imposed perturbation grows exponentially, in the linear regime of the evolution, and the system displays self-organized patterns at the non-linear stage of the evolution. Stationary stable patterns develop when the instability takes place on ordinary continuum space or on a symmetric graph. These are the celebrated Turing patterns, that typify on networks as a material segregation in activator rich and activator poor groups.  For reaction-diffusion systems on directed supports, waves are instead obtained as the late time echo of the instability.

Starting from these premises, we here wish to address the full non linear dynamics that stems for a topology driven instability, and eventually obtain a close form solution for the emerging traveling waves. To reach this goal we shall initialize the system right at the threshold of the instability ($\boldsymbol{ \mu} \equiv \boldsymbol{ \mu}_0$), when the real part of the dispersion relation is about to cross the horizontal axis, and then perturb the reaction parameter $\boldsymbol{ \mu}_0$ so as to make the homogeneous fixed point slightly unstable. A multiple time scale perturbative analysis, which accommodates for key topological ingredients, will open up the avenue to a detailed characterization of the complete non linear picture. 

When $\boldsymbol{ \mu} = \boldsymbol{ \mu} _0$,  the maximum value of $\lambda^{(\alpha)}_{Re}$ is therefore identically equal to zero, for a critical index $\alpha=\alpha_c$, to which corresponds a selected Laplacian eigenvalue  $\Lambda^{(\alpha_c)}=  \Lambda^{(\alpha_c)}_{Re}+i \Lambda^{(\alpha_c)}_{Im}$. Since $\Lambda^{(\alpha_c)}_{Im} \ne 0$, it follows \cite{asllani14} that $\lambda^{(\alpha_c)} \ne 0$. Indeed,  $\lambda^{(\alpha_c)} = \pm i \omega_0$, where $\omega_0= h(\Lambda_{Re}^{(\alpha_c)}) \Lambda_{Im}^{(\alpha_c)}$ with $h(\Lambda_{Re}^{(\alpha_c)}) = (2 D_x D_y \Lambda_{Re}^{(\alpha_c)}+ f_x D_y + g_y D_x)/\left [ f_x + g_y + ( D_y+ D_x) \Lambda_{Re}^{(\alpha_c)} \right ]$, as determined from a straightforward calculation.  From equation \eqref{eq:system2}, one can readily obtain  $\eta^{(\alpha_c)}= -(f_x + D_x \Lambda_{Re}^{(\alpha_c)})/f_y + i (\omega_0- D_x\Lambda_{Im}^{(\alpha_c)})/f_y$. The solution of the linear problem:
\begin{equation}\label{eq:first_u}
\left ( \frac{\partial }{\partial t} \mathbb{I}_{2N}    -{\bf L} - {\bf D} \right ) {\bf u} =0 
\end{equation}
is hence given by
\begin{equation}
\label{eq:sol_lin_ord}
{\bf u=U}_0 e ^{ i \omega_0 t} + c.c.
\end{equation}
where $c.c.$ stands for the complex conjugate. Here ${ \bf U}_0= \left (  \boldsymbol{ \phi}^{(\alpha_c)}  \; \; ;  \;  \eta^{(\alpha_c)}  \boldsymbol{ \phi}^{(\alpha_c)}  \right )$ is the right eigenvector of ${\bf L} + {\bf D}$ corresponding  to the eigenvalue $i \omega_0$. As we shall see, ${ \bf U}_0$ encodes the spatial characteristics of the predicted pattern.
\subsection{Multiscale  analysis: a topology dependent Stuart Landau equation}
Let us start from the neutral condition highlighted above, when the parameters are set to the marginal value $\boldsymbol{ \mu}_0$ that yields $\lambda^{(\alpha_c)}_{Re}=0$. Imagine to impose an appropriate perturbation in the form $\boldsymbol{ \mu}=\boldsymbol{ \mu}_0+ \epsilon^2 \boldsymbol{ \mu}_1$, where $\epsilon$ plays the role of a small parameter, and $\boldsymbol{ \mu}_1$ is order one. This modulation endows a tiny instability to develop: the dispersion relation acquires therefore a positive real part, which consistently scales as $\epsilon^2$. This latter observation sets the characteristic time scale for the examined instability, and opens up the perspective for a formal mathematical investigation. Following the prescription of the multiple time scale technique, we introduce $\tau= \epsilon^2 t$, the slow time variable, which we treat as independent from  time $t$. In the solution of the perturbation problem, the additional freedom introduced by the new independent time variable will be exploited to remove undesired secular terms. As we shall see, the latter set constraints on the approximate solution, which are called solvability conditions.

The total derivative with respect to the original time $t$ rewrites: 
\begin{equation}\label{eq:time}
\frac{d}{d t} \longrightarrow \frac{\partial}{\partial t} + \epsilon^2 \frac{\partial}{\partial \tau} \quad .
\end{equation}
Moreover, one may assume the following expansions to hold
\begin{equation}\label{eq:expansions}
\begin{aligned}
\bf
L&=& {\bf L}_0 + \epsilon^2 {\bf L}_1+\ldots \\
\mathcal{M}&=& \mathcal{M}_0 + \epsilon^2 \mathcal{M}_1+\ldots \\
\mathcal{N}&=& N_0 + \epsilon^2 \mathcal{N}_1+\ldots
\end{aligned}
\end{equation}
the unperturbed parameters $\boldsymbol{ \mu} _0$, and the associated correction factors $\boldsymbol{ \mu} _1$, being implicitly contained in the definition of the above operators. We further assume that ${\bf u}$, the solution of the non linear equation \eqref{eq:equation_u}, can be expressed as a perturbative series, function of both $t$ and $\tau$:
\begin{equation}\label{eq:u_pert}
{\bf u}(t) = \epsilon {\bf u}_1(t,\tau)+\epsilon^2 {\bf u}_2(t,\tau)+ \ldots
\end{equation}

To proceed in the analysis, one inserts equations \eqref{eq:time}, \eqref{eq:expansions} and \eqref{eq:u_pert} into  \eqref{eq:equation_u} to get:  
\begin{multline*}
\hspace{-0.25cm}\left(\frac{\partial}{\partial t  } \mathbb{I}_{2N} +\epsilon^2 \frac{\partial }{\partial \tau}\mathbb{I}_{2N}  - {\bf L}_0 -{ \bf D} -\epsilon^2 {\bf L} _1 - \ldots  \right ) (\epsilon { \bf u} _1 +\epsilon ^2  {\bf u}_2 +\ldots) \\
= \epsilon^2 \mathcal{M}_0 {\bf u}_1 { \bf u}_1 +\epsilon^3 (2 \mathcal{M}_0 { \bf u}_1 {\bf u}_2 +\mathcal{N}_0 {\bf u}_1 {\bf u}_1 {\bf u}_1) + \mathcal{O} (\epsilon^4) 
\end{multline*}
Equating terms of the same order in $\epsilon$ returns the following family of equations 
\begin{equation}\label{eq:terms}
\left ( \frac{\partial}{\partial t }  \mathbb{I}_{2N} - {\bf L}_0 -{\bf D}   \right ) {\bf u}_{\nu} = {\bf B}_{\nu} 
\end{equation}
with $\nu=1,2,3...$. Following the Fredholm theorem (see Appendix), the linear system \eqref{eq:terms} admits a non trivial solution if the solvability condition is satisfied, namely if $\langle ({\bf U}_0^* )^{\dagger} ,  {\bf B}_{\nu}^{(1)}\rangle=0$, where the angular brackets denotes the scalar product. 

We shall hereafter focus on the first three equations of the above hierarchy. The corresponding right-hand sides (see also Appendix) respectively read ${\bf B}_1=0$, $ {\bf B}_2=M_0 {\bf u}_1 {\bf u}_1$ and ${\bf B}_3= (-\frac{\partial}{\partial \tau} \mathbb{I}_{2N} +{\bf L}_1) {\bf u}_1 + 2 \mathcal{M}_0 {\bf u}_1 {\bf u}_2 +\mathcal{N}_0 {\bf u}_1 {\bf u}_1 {\bf u}_1 $. The solvability condition is naturally met for $\nu=1,2$, while it needs to be explicitly imposed for $\nu=3$. 

Consider first the leading order contribution, $\nu=1$ and solve the corresponding differential equation for ${\bf u}_1$. As expected, this is equivalent to equation \eqref{eq:first_u}, that we derived under the linear approximation. Hence, ${\bf u}_1$ follows from \eqref{eq:sol_lin_ord} modified with the inclusion of an arbitrary, complex and so far undermined, amplitude factor  $ W(\tau)$, function of the slow time scale $\tau$. In formulae:    
\begin{equation}
\label{eq:u1_W} 
{\bf u}_1(t, \tau) = W(\tau) {\bf u} =  W(\tau) {\bf U}_0 e^{ i \omega_0 t} + c. c. 
\end{equation}

As we will see, the factor $ W(\tau)$ sets the typical amplitude of the emerging patterns: it should be constrained to match the required solvability condition and so self-consistently determined via the multiple scale calculation. As already emphasized, equation \eqref{eq:u1_W} constitutes a natural generalization of the linear solution \eqref{eq:sol_lin_ord}, which indirectly accommodates for the non-linearities through the slow varying amplitude factor $W$. This will in turn enable us to track the time evolution of the patterns, beyond the initial stage of the exponential growth. The remaining part of the calculation is devoted to deriving a consistent equation for the time evolution of the complex amplitude $W$. As we shall see, this amounts to imposing the solvability condition at $\nu=3$.

To solve the next-to-leading order ($\nu=2$) equation in \eqref{eq:terms}, we put forward the following ansatz \cite{kuramotoBook} for ${\bf u}_2$:  
\begin{equation*}
{\bf u}_2=  {\bf V}_+  e^{2 i \omega_0 t } + {\bf V}_- e^{-2 i \omega_0 t } + v_0 {\bf u}_1 + {\bf V}_0 
\end{equation*}
The constant $v_0$ cannot be determined at this stage, and will not affect the forthcoming developments. Inserting in \eqref{eq:terms} and grouping together the terms that do not depend on $t$, one finds $ {\bf V}_0 = -2 \vert W \vert ^2 \left ( {\bf L}_0 + {\bf D} \right) ^{-1} \mathcal{M}_0 {\bf U}_0 \bar{{\bf U}} _0$ where the bar stands for the conjugate.  Similarly, equating the terms proportional to $e^{2 i \omega_0 t } $ (resp. $e^{-2 i \omega_0 t } $)  yields   $ {\bf V}_{+} = {\bf V}_{-}= W^2 \left ( 2 i \omega_0 \mathbb{I}_{N} -L_0 -D \right ) ^{-1} \mathcal{M}_0 {\bf U}_0 {\bf U}_0$. 

At the next order in the hierarchy, $\nu=3$, the linear equation for ${\bf u}_3$ builds on the above characterization for both ${\bf u}_1$ and ${\bf u}_2$. In particular, the unknown complex amplitude $W$ enters the definition of the right-hand side ${\bf B}_3$. By imposing the solvability condition $\langle ({\bf U}_0^* )^{\dagger} ,  {\bf B}_{3}^{(1)}\rangle=0$, and carrying out a straightforward manipulation, one eventually obtains the following SL equation for $W(\tau)$ 
\begin{equation}\label{eq:SLE}
\frac{d}{d \tau} W(\tau) = g^{(0)} W -  g^{(1)} \vert W \vert ^2 W
\end{equation}
where $ g^{(0)} \equiv g^{(0)} _{Re} + i g^{(0)} _{Im}= ({\bf U}_0^*)^{\dagger} {\bf L}_1 {\bf U}_0 $ and $ g^{(1)} \equiv g^{(1)} _{Re} + i g^{(1)} _{Im}=-({\bf U}_0^*)^{\dagger} \left [ 2 \mathcal{M}_0 {\bf V}_+ \bar{\bf U}_0 \right . $ $ \left .  +2 \mathcal{M}_0   {\bf V}_0 {\bf U}_0 + 3 \mathcal{N}_0  {\bf U}_0   {\bf U}_0    \bar{\bf U}_0   \right ]$ are complex numbers. Notice that $g^{(0)}$ and $g^{(1)}$ depend both on the reaction terms of the original  system \eqref{eq:systemMF}, through  e.g. ${\bf L}_1$, $\mathcal{M}_0$, $\mathcal{N}_0$, and on the topological characteristics of the embedding network, via ${\bf U}_0$. To derive equation \eqref{eq:SLE} use has been made of the normalization condition $( {\bf U}_0^* )^{\dagger} {\bf U}_0^* = 1$.

The solution of equation \eqref{eq:SLE} can be cast in the form:  
\begin{equation}\label{eq:solutionSL}
W(\tau)= \sqrt{\frac{ g^{(0)} _{Re}}{\vert  g^{(1)}_{Re } \vert}}  \: \exp \left [ i \left (g^{(0)}_{Im} -  g^{(1)}_{Im}  \frac{ g^{(0)} _{Re}}{\vert  g^{(1)}_{Re } \vert} \right ) \tau  + i  \psi \right] 
\end{equation}
where $\psi$ is a phase term which relates to the assigned initial conditions.

Summing up, and recalling equation \eqref{eq:u1_W}, the wave-like pattern $\boldsymbol{ \mathsf{x}}(t, \epsilon)$, instigated by the directed network, close to the threshold of instability, will be approximately described by:
\begin{multline}\label{eq:solutionSL_final}
\boldsymbol{ \mathsf{x}}(t, \epsilon) =  \boldsymbol{ \mathsf{x^*}}+\\
\left(\epsilon \sqrt{\frac{ g^{(0)} _{Re}}{\vert  g^{(1)}_{Re } \vert}}
 {\bf U}_0 \exp \left[i \omega_0 t+ i \left (g^{(0)}_{Im} -  g^{(1)}_{Im}  \frac{ g^{(0)} _{Re}}{\vert  g^{(1)}_{Re } \vert} \right ) \epsilon^2 t  \right] +c.c. \right) 
\end{multline}
where we have arbitrarily set $\psi=0$.  As anticipated, the structure of the graph which ultimately drives the instability enters parametrically the above solution \eqref{eq:solutionSL_final}. In the following, we shall indicate with $A_{x,y}$ the amplitude of the oscillating patterns, for respectively species $x$ and $y$, around the average solution. 
\subsection{Alternative perturbation scheme: acting on the diffusion coefficients}
In the previous section we have seen how to characterize the emerging patterns when they originate from a perturbation of the reaction coefficients $\boldsymbol{ \mu} $. Similarly, one could imagine to induce the instability by perturbing the diffusion constants from $D_x$ and $D_y$. More specifically, we initialize the unperturbed system so as  to match the marginal condition $\lambda^{(\alpha_c)}_{Re}=0$ and then perform the change $D_x \rightarrow D_x + \epsilon^2 D_{x1}$ and $D_y\rightarrow D_y + \epsilon^2 D_{y1}$, where $\epsilon$ is a small parameter, and  $D_{x1}$  and  $D_{y1}$ are order one scalar quantities. Proceeding in analogy with the above yields a SL differential equation for the evolution of the complex amplitude factor $W(\tau)$, where  $g_1$ is unchanged and $g_0=({\bf U}_0^*)^{\dagger} {\bf D}_1 {\bf U}_0 $ where 
\begin{equation*}
{ \bf D_1 } = \left (
\begin{array}{c|c}
D_{x1}  {\bf  \Delta }  & \mathbb{O}_{N} \\
\phantom{i } & \phantom{i }\\
\hline
\phantom{i } & \phantom{i }\\
\mathbb{O}_{N}  & D_{y1} {\bf  \Delta }
\end{array}
\right)  \qquad .
\end{equation*}
\subsection{Numerical validation of the theory}
We here aim at testing the predictions of the theory, by drawing a comparison with the outcome of direct simulations performed for a reaction-diffusion model of paradigmatic interest. This is the celebrated Brusselator model, a non linear reaction scheme which describes the autocatalytic coupling of two mutually interacting chemical species.  Details of this model can be found in the Appendix. 
\begin{figure*}[h]
\begin{center}
\begin{tabular}{ccc}
\includegraphics[scale=0.206]{relDisp_B_reteNW.eps}&
\includegraphics[scale=0.3]{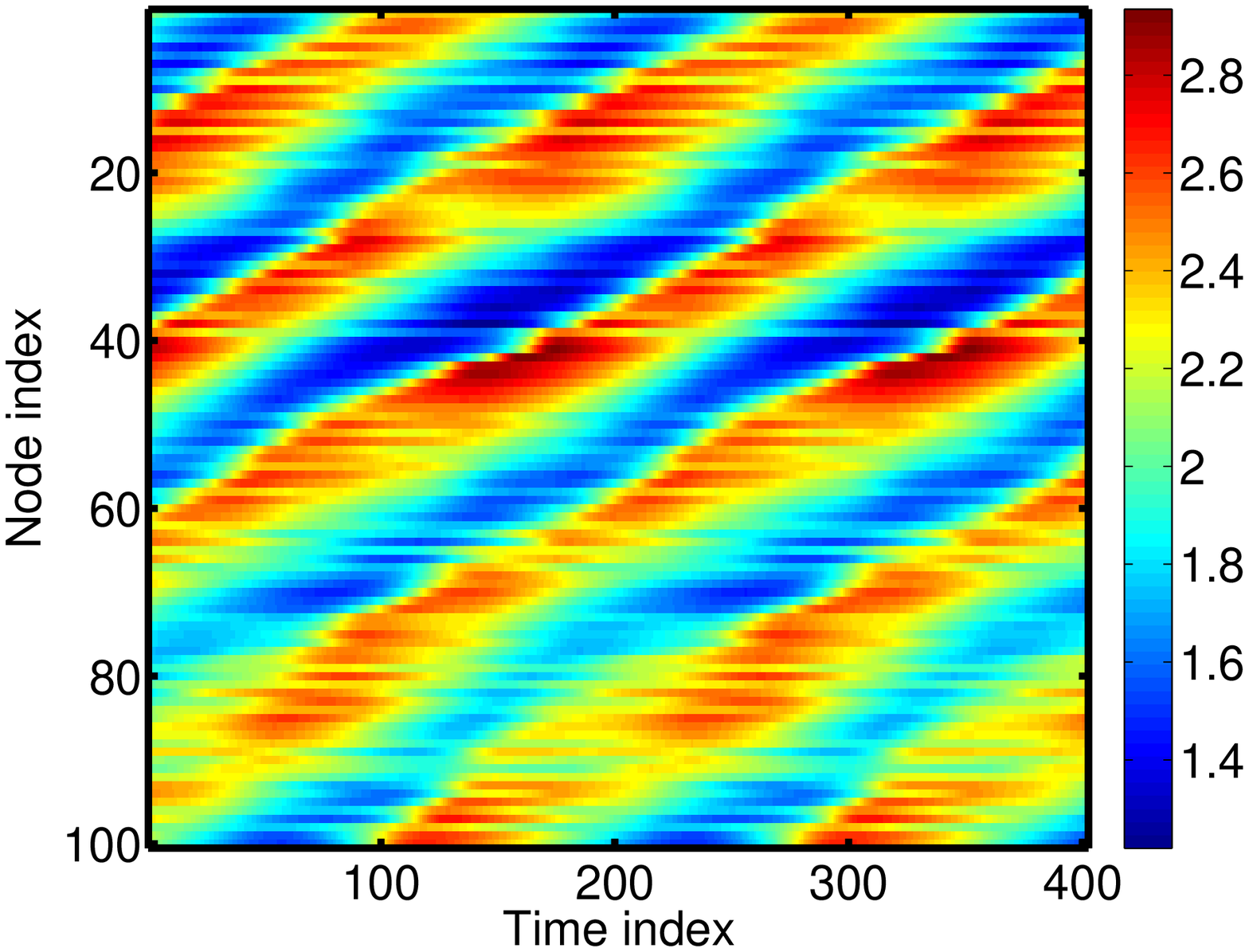}&
\includegraphics[scale=0.3]{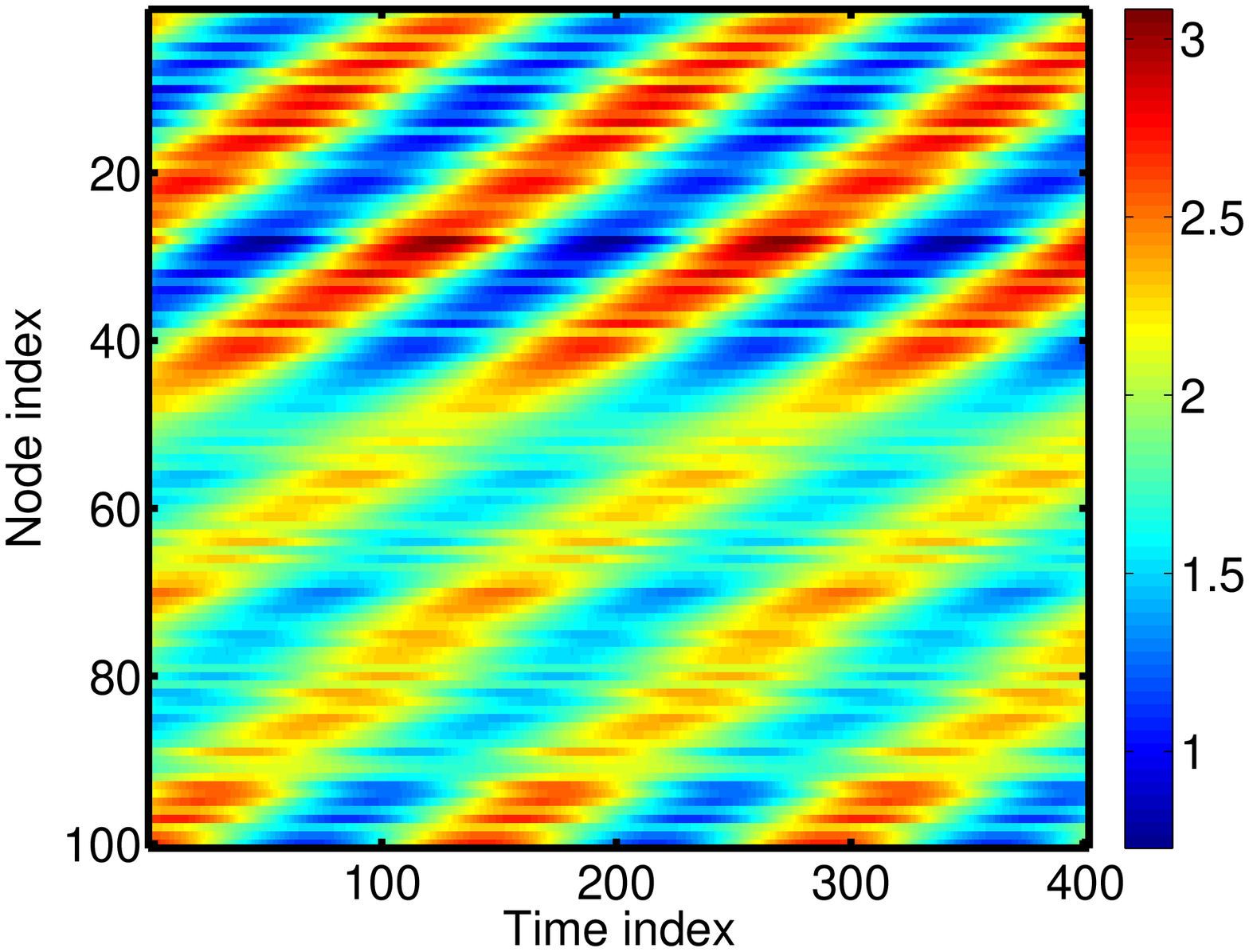} \\
(a) & (b) & (c)
\end{tabular}
\end{center}
\caption{Panel (a): The (blue, online) triangles represent the real part of the dispersion relation as a function of (minus) the real part of the eigenvalues of the hosting directed network. The black line originates from the continuous theory. Panel (b): Pattern emerging from  species $y$ as obtained by direct integration of system \eqref{eq:systemMF}. The concentration on each node is plotted as a function of time. Panel (c): Pattern relative to species $y$ as determined from the analytical solution of \eqref{eq:solutionSL_final}. The network is made of $N=100$ nodes and has been generated following the NW recipe with $p=0.27$.   Parameters are $a=2.1$, $b_c=4.002$, $c=1$, $d=1$, $D_x=1$ and $D_y=3$.  The perturbation is here acting on $b$ as $b=b_c(1+\epsilon^2)$, with $\epsilon^2=0.1$. \label{fig:reteNW}}
\end{figure*}

As a first example, we consider the Brusselator model defined on a balanced network generated with a slightly modified version of the Newman-Watts (NM) algorithm \cite{newman99} (see Appendix). In the left panel of Figure \ref{fig:reteNW}  we display with symbols the real part of the dispersion relation  $\lambda^{(\alpha)}_{Re}$ as a function of the real part of the Laplacian eigenvalue $\Lambda^{(\alpha)}_{Re}$ (changed in sign). The parameters of the model have been set so  as to have the largest value of $\lambda^{(\alpha)}_{Re}$ equal to zero, in correspondence of a specific  $-\Lambda^{(\alpha_c)}_{Re}$. The solid line represents instead the dispersion relation obtained, with the same choice of the parameters, for the limiting case of a symmetric continuous support. If the system is placed on top of a symmetric graph, the continuous curve turns into a discrete collection of points, following exactly the same profile and reflecting the finite set of (real) eigenvalues, associated to the Laplacian operator. When the embedding network is instead asymmetric, the complex component of the Laplacian spectrum lifts the dispersion relation, as depicted in leftmost panel of Figure  \ref{fig:reteNW}, so eventually inducing a topology driven instability, in a otherwise dynamically stable system. In the other two panels of Figure  \ref{fig:reteNW} the patterns obtained via a numerical integration of the reaction-diffusion system  \eqref{eq:systemMF} and the analytical solution \eqref{eq:solutionSL_final} are respectively reported, displaying a satisfying degree of correspondence.
\begin{figure}[tb]
\begin{center}
\includegraphics[scale=0.3]{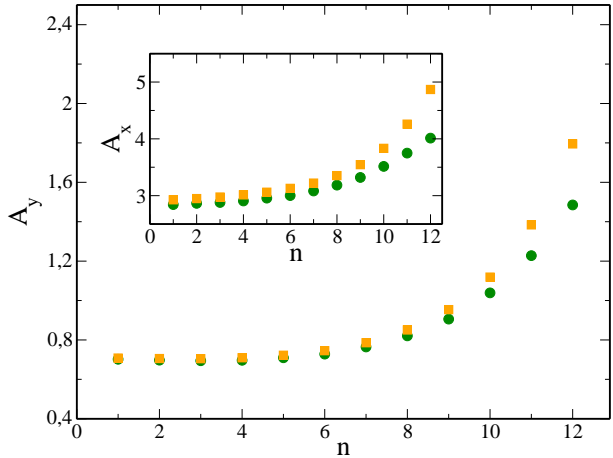}
\end{center}
\caption{Amplitude of the self-emerging oscillations as a function of the shift parameter $n$. Here $K=27$ and $n \in (0,12)$.  Orange dots refer to the amplitudes obtained from equation \eqref{eq:solutionSL_final}, while green symbols follow from numerical integration of the Brusselator model. The main panel refers to the amplitude of the patterns relative to
species $y$, while, in the inset, the amplitudes are calculated for species $x$.  Parameters are $a=4$, $c=1$, $d=1$, $D_x=1$, $D_y=3$. The instabilities come from the perturbation of parameter $b$ as $b=b_c(1+\epsilon^2)$ for $\epsilon^2 = 0.1$. For each $n$, the critical value of $b_c$ is calculated so as to satisfy the condition $max(\lambda_{Re}^{(\alpha)})=\lambda_{Re}^{(\alpha_c)}=0$ . \label{fig:ampiezzaB}}
\end{figure}

\begin{figure*}[t]
\begin{center}
\begin{tabular}{ccc}
\includegraphics[scale=0.3]{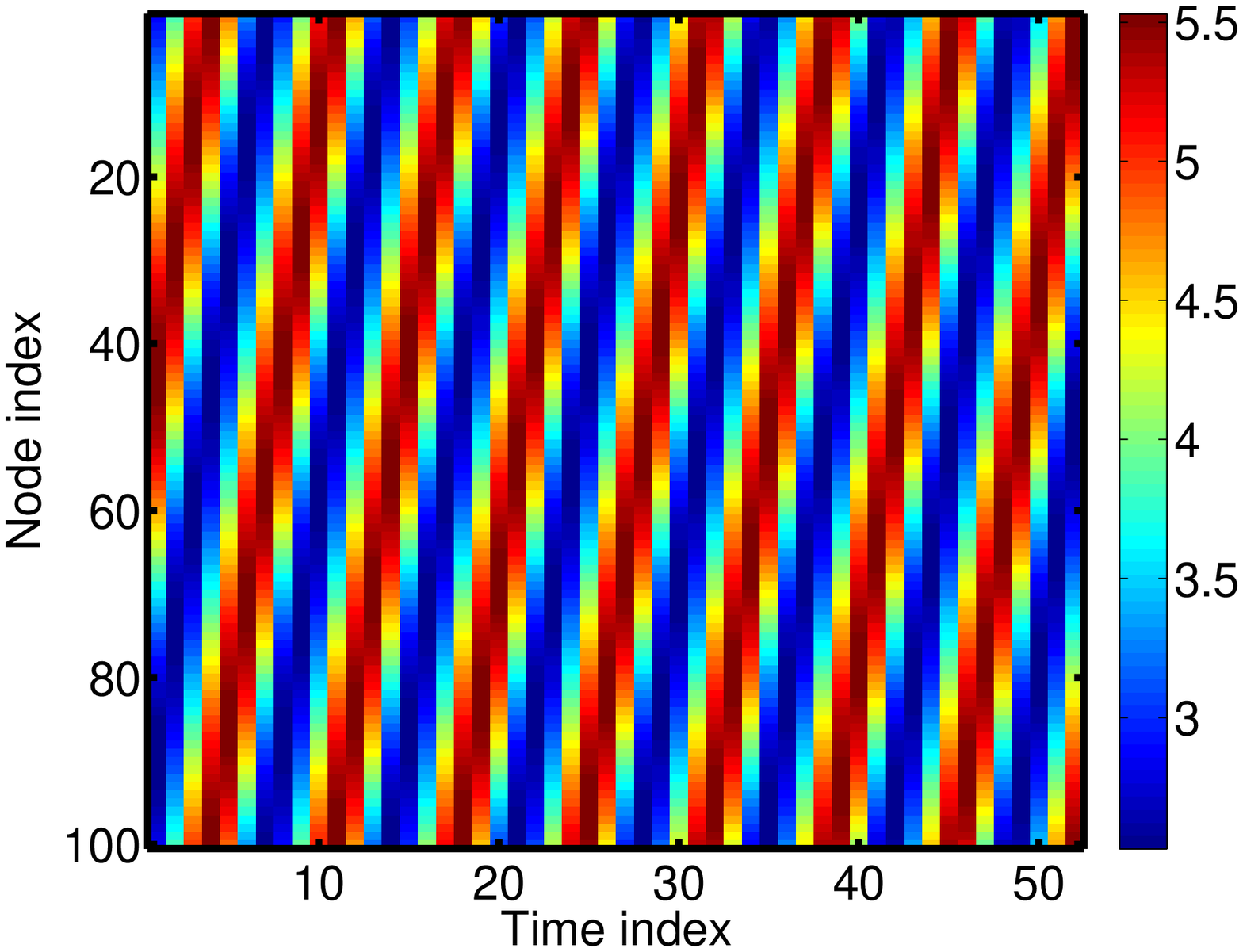} &
\includegraphics[scale=0.3]{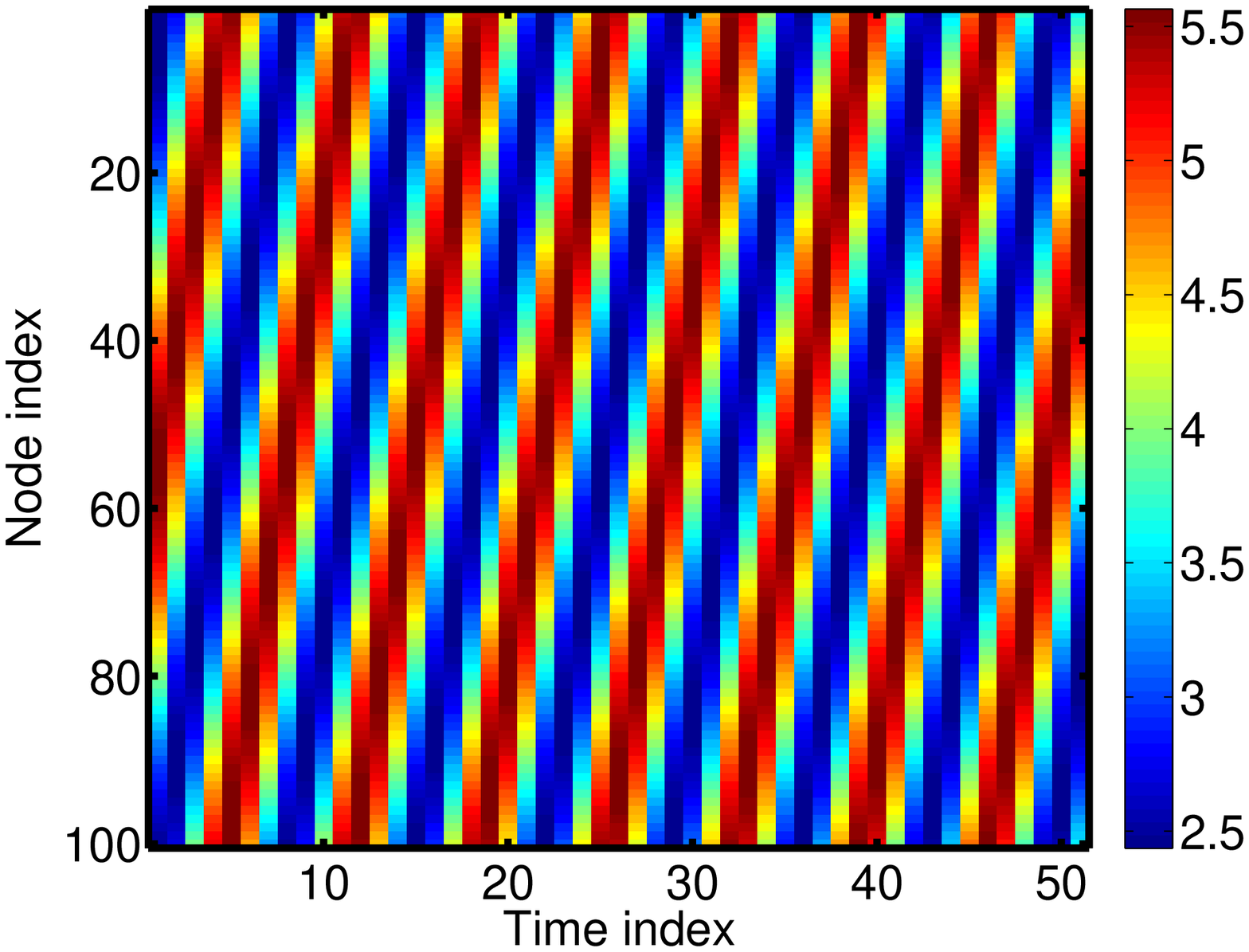}&
\includegraphics[scale=0.206]{sincronizzatiB.eps}\\
(a) & (b) & (c)
\end{tabular}
\end{center}
\caption{Waves on a directed lattice obtained by imposing $n=5$ shifts. For the parameters' description refer to the caption of Figure \ref{fig:ampiezzaB}. Panels (a) and (b) show the time dependent patterns relative to species $x$ obtained, respectively, from the original Brusselator model and the SL equation. In panel (c) we display $x$ versus the index of the node, while the inset reports $x$ as a function of time $t$, for a selected node. In both cases, green circles refer to data extracted from panel (a) (simulations), orange squares to panel (b) (theory). \label{fig:soluzioniB}}
\end{figure*}

As an additional check for the developed theory, we consider a family of directed regular lattices, with varying level of imposed asymmetry. More specifically, we preliminary assumed a closed one dimensional ring composed of $N$ nodes: each node has $K$ links to its first $K$ nearest neighbors encountered when circulating the ring clockwise. The adjacency matrix which describes such a lattice is then shifted, via $n$ successive applications of a one-step shift operator, so to result in a set of distinct lattices, which tend to progressively approach the symmetric limiting case. For such $n$ directed networks, we computed the amplitude of the predicted, topology driven patterns, as follows equation \eqref{eq:solutionSL_final}, and compared it to the outcome of numerical simulations based on the original reaction-diffusion model. Results of the analysis are reported in Figure \ref{fig:ampiezzaB}, where the amplitude of the pattern is plotted as a function of the degree of shift $n$. Here, the instability is produced upon perturbation of the reaction parameter $b$. An overall excellent agreement is observed, between theory and simulations. The predictive adequacy of the theory can be also appreciated in Figure \ref{fig:soluzioniB} where the time dependent patterns are displayed for $n=5$, a representative case study.  The same conclusion holds when the perturbation acts on the diffusion coefficient $D_x$ and $D_y$ (data not shown). 

%
%
%
%
%
%
\section{Discussion}
Self-organized patterns can spontaneously develop in a multi-species reaction-diffusion system, as follow a symmetry breaking instability of an homogeneous equilibrium. Inhomogeneous perturbation can in fact amplify due to the constructive interference between reaction and diffusion terms, and eventually yield coherent, spatially extended motifs in the non-linear regime of the evolution. Reaction-diffusion systems placed on symmetric graphs have been also analyzed in the literature. The conditions for the deterministic instability are derived via a linear stability analysis, which requires expanding the perturbation on a complete basis formed by the eigenvectors of the discrete Laplacian. For system hosted on undirected networks, the instability is essentially driven by nonlinearities, which stem from both reactions and diffusion. The topology of the embedding network-like support defines the relevant directions for the spreading of the perturbation, but cannot influence the onset of the instability. A radically different scenario is encountered when a directed graph is instead assumed to provide the spatial backing for the scrutinized model. In this case, the topology of the space is equally important and significantly impact the conditions that drive the dynamical instability.

Building on these recent advances, the aim of this paper is to go beyond the standard linear stability analysis for the outbreak of the instability and provide a complete characterization of the patterns emerging on a directed discrete support, in the fully developed non linear regime. To this end we have applied a multiple time scale analysis, generalizing to the present context the original derivation of \cite{kuramoto75}. This results in a cumbersome calculation owing to the particular nature of the diffusive coupling imposed. The amplitude of the most unstable mode is shown to obey a Stuart-Landau (SL) equation whose coefficients unprecedently reflect the topology of the network, the genuine drive to the instability. Simulations performed for the Brussellator model, confirm the validity of the theory, which proves effective in quantitatively grasping the characteristics of self-emerging dynamical patterns, close to the threshold of instability. This is a significant achievement which could translate in novel strategies to control the dynamics of the system, via appropriate fixing of topological features, including the supervised addition/removal of specific nodes/links in the network.  
\section{Appendix}
\subsection{The solvability condition} 
Let $\bf A$ be a linear operator, and ${\bf u}(t)$ and ${\bf b} (t)$ two complex vectors of the same length.  According to the Fredholm theorem, a linear system ${ \bf A u}(t)={\bf b}(t)$ is solvable if $ \langle {\bf v} (t), {\bf b} (t) \rangle=0$ for all vectors ${\bf v}(t)$ solution of  ${\bf A}^{*} {\bf v}(t) =0$, where ${\bf A}^{*}$ is the adjoint operator satisfying  $\langle {\bf A} ^* \bf y, x \rangle = \langle  y,  A x \rangle$ $\forall x,y$. The angular brackets denote the scalar product that we here define as $\langle { \bf v}(t), {\bf b}(t) \rangle $ $ =\int_0^{2 \pi / \omega_0} {\bf v}^{\dagger}(t) {\bf b}(t) dt$, the symbol $\dagger$ standing for the conjugate transpose. With reference to equation \eqref{eq:terms}, the first requirement of the Fredholm theorem consists in finding ${\bf v}(t)$ such that  $(\partial / \partial t \mathbb{I}_{2 N} -{\bf L}_0 - {\bf D})^{*} {\bf v}(t)=0$. Recalling that ${\bf L}_0$ and $\bf D$ are real matrices, by partial integration we find that $(\partial / \partial t \mathbb{I}_{2 N} -{\bf L}_0 - {\bf D})^{*} =-(\partial / \partial t \mathbb{I}_{2 N} +{\bf L}_0 +{\bf D})^{T}$. As a consequence, the system to be solved is $-(\partial / \partial t \mathbb{I}_{2 N} +{\bf L}_0 +{\bf D})^{T} {\bf v}(t)=0$. In analogy with equation \eqref{eq:first_u}, we search ${\bf v}(t)$ in the form ${\bf v} (t)={\bf U}_0^* e^{i \omega_0 t}$ for some vector ${\bf U}_0^*$. Substituting this ansatz into the previous equation, we find  $({\bf L}_0+{\bf D})^T {\bf U}_0^*=-i \omega_0 {\bf U}_0^*$. In analogy with ${\bf U}_0$, ${\bf U}_0^*$ is related to the eigenvalue problem $\boldsymbol{\Delta}^{T} \boldsymbol{\psi}^{(\alpha_c)} =(\Lambda_{Re}^{\alpha_c}- i \Lambda_{Im}^{\alpha_c} ) \boldsymbol{\psi}^{(\alpha_c)}$ through ${ \bf U}_0^*= \left (  \boldsymbol{ \psi}^{(\alpha_c)}  \quad  \eta^{(\alpha_c)}_*  \boldsymbol{ \psi}^{(\alpha_c)_0}  \right )^T$ with $\eta^{(\alpha_c)}_*= -(f_x + D_x \Lambda_{Re}^{(\alpha_c)})/g_y - i (\omega_0- D_x\Lambda_{Im}^{(\alpha_c)})/g_y$. Having defined ${\bf U}_0^*$, one can explicitly write the  solvability condition $\langle {\bf U}_0^* e^{i \omega_0 t} , {\bf B}_{\nu} (t, \tau) \rangle =0$. Since $ {\bf B}_{\nu} (t, \tau)$ turns out to be periodic functions of period $2 \pi /\omega_0$, it is appropriate to express them in the form ${\bf B}_{\nu} (t, \tau) =\sum_{l=-\infty}^{+\infty} {\bf B}_{\nu}^{(l)} (\tau) e^{i l \omega_0 t}$. If we multiply this series by  $({\bf U}_0^* e^{i \omega_0 t})^{\dagger}$ we again obtain periodic functions that, when integrated over the period $2\pi /\omega_0$ give zero. The only exception holds for $l=1$ which gives $\langle {\bf U}_0^* e^{i \omega_0 t} ,  {\bf B}_{\nu}^{(1)} (\tau) e^{i  \omega_0 t} \rangle = \int_0^{2 \pi / \omega_0} ({\bf U}_0^*)^{\dagger}  {\bf B}_{\nu}^{(1)} (\tau) dt$. The integrand does not depend on time $t$ and therefore the integral is zero only if the integrand itself  is identically equal to zero. For this reason the solvability condition reduces to $({\bf U}_0^* )^{\dagger}  {\bf B}_{\nu}^{(1)} (\tau)=0$ $\forall \nu$.

\subsection{The Brusselator model} 
In the Brusselator model, the two reaction terms are specified by $f(x_i, y_i, {\bf \mu})=a-(b+d) x_i+c x_i^2 y_i$ and $g(x_i, y_i, {\bf \mu})=b x_i -c x_i^2 y_i$, where ${\bf \mu}=(a, b, c, d)$ defines a set of positive real parameters. The unique homogeneous equilibrium point is  $(x^*, y^*)=(a/d, bd/c/a)$.

\subsection{Network generation strategy}
We start from a substrate $K$-regular ring made of $N$ nodes. The NW algorithm \cite{newman99} is designed to add, on average, $NKp$ long-range directed links, in addition to the links that originate from the underlying regular lattice. Here $p$ lies in the interval $[0, 1]$ and represents a probability to be chosen by the user. The NW algorithm here employed is modified so as to result in a balanced network (identical number of incoming and outgoing links, per node). To this end, the inclusion of a long-range link starting from node $i$ is accompanied by the insertion of a fixed number ($3$ is our arbitrary choice) of additional links to form a loop that closes on $i$.
%
%
%
%
%
%
%
%
%
%
\acknowledgments
This work has been partially supported by Ente Cassa di Risparmio di Firenze and program PRIN 2012 founded by the Italian Ministero dell’Istruzione, dell'Universit\`{a} e della Ricerca (MIUR).

%
%
%
%
%
%

\bibliography{bibliography}{}
\bibliographystyle{plain}
%
%
%
%
%
%
\end{document}